\documentclass[conference]{IEEEtran}
\IEEEoverridecommandlockouts
\usepackage{cite}
\usepackage{amsmath,amssymb,amsfonts}
\usepackage{graphicx}
\usepackage{amsmath}
\usepackage{textcomp}
\usepackage{subcaption}
\usepackage{algorithm}
\usepackage{float}
\usepackage{algpseudocode}
\usepackage[margin=0.8in, top=0.8in,bottom=0.9in, columnsep=0.24in]{geometry}
\usepackage{array}
\usepackage{xcolor}
\usepackage{caption}
\def\BibTeX{{\rm B\kern-.05em{\sc i\kern-.025em b}\kern-.08em
    T\kern-.1667em\lower.7ex\hbox{E}\kern-.125emX}}
\begin{document}

\title{On the Application of Deep Learning for Precise Indoor Positioning in 6G}

\author{\IEEEauthorblockN{Sai Prasanth Kotturi\IEEEauthorrefmark{1},
Anil Kumar Yerrapragada\IEEEauthorrefmark{2},
Sai Prasad\IEEEauthorrefmark{3},
Radha Krishna Ganti\IEEEauthorrefmark{4}}
\IEEEauthorblockA{Department of Electrical Engineering, Indian Institute of Technology Madras, Chennai-600036, India.\\
Email: \IEEEauthorrefmark{1}ksaiprasanth7@smail.iitm.ac.in,
        \IEEEauthorrefmark{2}anilkumar@5gtbiitm.in,
        \IEEEauthorrefmark{3}venkatasiva@5gtbiitm.in,
	\IEEEauthorrefmark{4}rganti@ee.iitm.ac.in
}
}
\maketitle

\begin{abstract}
Accurate localization in indoor environments is a challenge due to the Non Line of Sight (NLoS) nature of the signaling. In this paper, we explore the use of AI/ML techniques for positioning accuracy enhancement in Indoor Factory (InF) scenarios. The proposed neural network, which we term LocNet, is trained on measurements such as Channel Impulse Response (CIR) and Reference Signal Received Power (RSRP) from multiple Transmit Receive Points (TRPs). Simulation results show that when using measurements from 18 TRPs, LocNet achieves a 9 cm positioning accuracy at the 90th percentile. Additionally, we demonstrate that the same model generalizes effectively even when measurements from some TRPs randomly become unavailable. Lastly, we provide insights on the robustness of the trained model to the errors in ground truth labels used for training.
\end{abstract}


\section{Introduction} \label{intro}
Precise user positioning is critical for 6G applications such as self-driving cars, digital twins, augmented and virtual reality, and the Industrial Internet of Things (IIoT)~\cite{malakuti2018architectural}. Several stakeholders from industry, academia, and standards bodies~\cite{imt2030_framework} have recognized user positioning, both indoor and outdoor, as a fundamental technology. 

Conventional positioning methods are of two types - those that depend on signalling from Radio Access Technologies (RAT) such as cellular systems and those that require signalling from satellites such as the Global Navigation Satellite System GNSS). RAT-dependent positioning methods are based on (1) time measurements such as Time of Arrival (ToA) ~\cite{xiong2013arraytrack} and Time Difference of Arrival (TDoA) ~\cite{zhang2013tdoa}, (2) angle measurements such as Angle of Departure (AoD) ~\cite{schmidt1986multiple} and Angle of Arrival (AoA) ~\cite{li1993performance}, (3) hybrid measurements involving a combination of times and angles~\cite{singh2021high}, (4) power measurements~\cite{nikonowicz2022indoor} and (5) carrier phase measurements~\cite{nikonowicz2022indoor}. These methods, though adequate in scenarios dominated by LoS signal paths, do not achieve good positioning accuracies in NLoS dominant scenarios such as Indoor Factories. The performance degradation is mainly due to the fact that NLoS paths lead to inaccurate estimation of timing, angle, power and phase.

Proper estimation of parameters like ToA, TDoA and AoA relies on knowledge of channel characteristics either in the form of Channel Impulse Response (CIR), Power Delay Profile (PDP) or Delay Profile (DP). The estimation of such parameters is inherently non-linear, leading us to explore AI/ML-based methods. AI/ML can also be used to estimate the user position directly~\cite{mogyorosi2022positioning}, avoiding timing and angle measurements altogether. Depending on the output of the AI/ML model, there are two types of positioning methods. 

\subsubsection{Direct AI/ML}
\begin{figure}[H]
    \centering
    \includegraphics[width=0.485\textwidth]{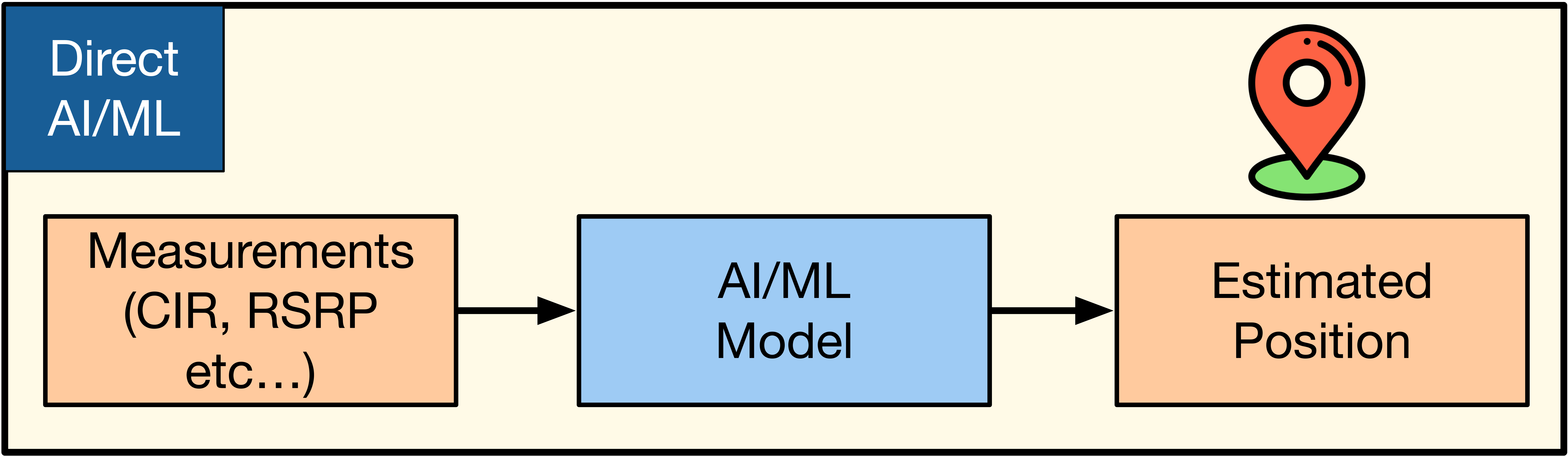}
    \caption{Direct AI/ML Positioning}
    \label{fig: direct_aiml}
\end{figure}

As shown in Figure~\ref{fig: direct_aiml}, these methods use the AI/ML model to directly estimate the position, taking measurements such as CIR, RSRP etc as inputs~\cite{SaiPrasanth}. 

\subsubsection{AI/ML Assisted}
\begin{figure}[H]
    \centering
   \includegraphics[width=0.485\textwidth]{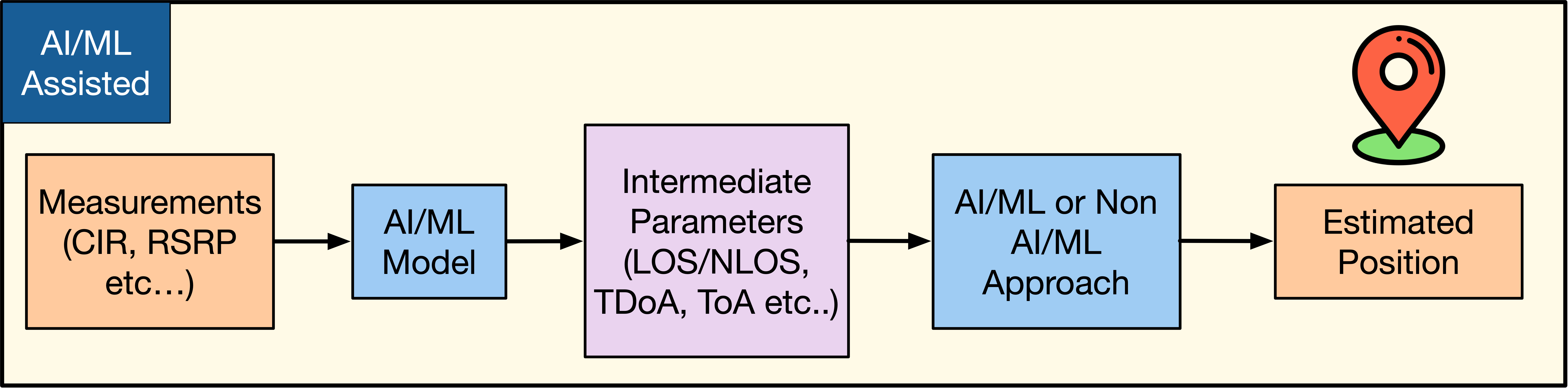}
    \caption{AI/ML Assisted Positioning}
    \label{fig: aiml_assisted}
\end{figure}

As shown in Figure~\ref{fig: aiml_assisted} these methods take measurements such as CIR and RSRP as model input to perform intermediate tasks such as LoS/NLoS classification and ToA prediction. These intermediate parameters are then used to determine the position, either through an AI/ML or non-AI/ML approach. 

\subsection{Our contributions}
This paper looks at the application of Direct AI/ML for positioning accuracy enhancements. We make the following contributions:
\begin{itemize}
\item We develop a neural network architecture that uses residual blocks and attention mechanisms. The model takes channel measurements from multiple TRPs as inputs, and predicts the $(x, y)$ position of a UE. The model is developed for one example of an NLoS dominant scenario, the Indoor Factory (InF). We compare the performance of the model with that of more conventional residual networks without attention mechanisms. 
\item In developing the aforementioned neural network, we answer the following 4 questions: 
    \begin{itemize}
        \item What is the best model input combination that achieves centimeter-level positioning accuracy?
        \item Can a single AI/ML model be used irrespective of the number of TRPs providing channel measurements?
        \item What is the minimum number of measurements that ensure centimeter-level positioning accuracy?
        \item What is the impact of ground truth label error on model performance?
    \end{itemize}
\end{itemize}

\section{Background}
Positioning methods have played an crucial role in 5G communication systems by supporting various applications ranging from navigation and location-based services to enhanced network management and emergency services. There are many positioning methods that have been explored extensively for both indoor and outdoor scenarios using GPS~\cite{zhou2024deep}, cellular communications~\cite{yang2024positioning} and Wi-Fi systems~\cite{10472472}. 

Specifically for the indoor environment, ~\cite{nikonowicz2022indoor} shows that Carrier Phase (CP) enhancements in InF-LoS scenarios can achieve centimeter-level positioning accuracy. However, the results show a significant drop in the positioning accuracy using CP in InF-NLoS environments with high clutter density. The carrier phase measurements are less reliable in InF-NLoS scenarios because of the high signal reflections that introduce phase errors, making it difficult to determine the exact number of full wavelengths between the transmitter and receiver (integer ambiguity). The RAT dependent positioning techniques used in the 5G NR releases, perform poorly in NLoS scenarios. With these methods, the $90$th percentile positioning accuracy is greater than $15m$ as shown in~\cite{3gpp_38_843}. 

In order to combat these challenges, AI/ML based cellular positioning for 6G applications is being studied~\cite{3gpp_38_843}. For NLoS scenarios, ~\cite{chatelier2023influence} uses deep learning to estimate the position. The work in~\cite{chatelier2023influence} studies the influence of various dataset parameters like the type of radio signals to estimate the position, number of TRPs and dataset size on the position accuracy. Two different types of radio data i.e., CIR and Path Gain (PG) are studied, and it was shown that the CIR data type provides better positioning accuracy with less number of TRPs compared to the PG data type. The positioning accuracy achieved in~\cite{chatelier2023influence}, is in the range of $2$ to $5$ meters. 

To achieve centimeter level accuracy, in this work we employ the following techniques:
\begin{itemize}
    \item Use of a combination of different radio measurements for model training
     \item Use of an Attention Mechanism to augment the ResNet-like architecture that is less complex than traditional ResNets and achieves higher position accuracy.
    \item AI/ML Model generalisation with respect to variable number of TRPs that provide measurements.
    \item Robustness of the model to the inherent errors in ground truth labels. 
\end{itemize}
In this work we propose a deep learning architecture called LocNet (Location Net) presented in Section~\ref{LocNet Model} to achieve sub-centimeter level accuracy. We use the LocNet model for all of our simulations.

\section{System Model}
This section describes the Indoor Factory layout, its channel model and the channel measurements that can be used as inputs for AI/ML based positioning. We consider the specific Indoor Factory scenario known as InF-DH~\cite{3gpp-38-901}. which is characterized by Dense clutter and High base stations (TRPs).

\subsection{InF-DH Scenario Layout} \label{InF-DH Layout}
\begin{figure}[H]
    \centering
    \includegraphics[width=0.485\textwidth]{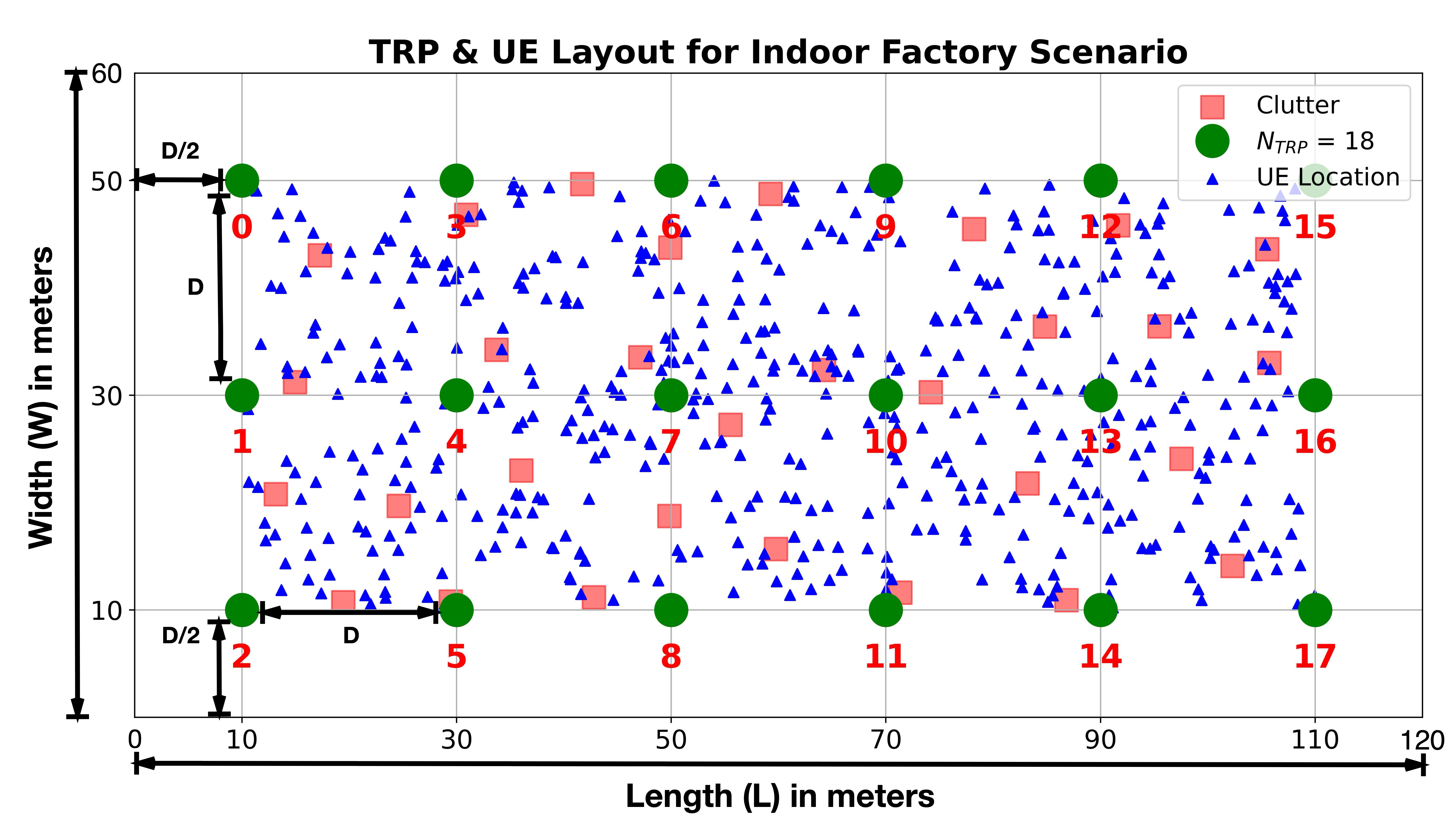}
    \caption{Example InF-DH scenario layout with $L = 120m$, $W = 60m$, $D = 20m$, and clutter ($r$, $h_c$, $d_{c}$) = $(60\%, 6m, 2m)$.}
    \label{fig:Example InF-DH Layout}
\end{figure}

 The layout consists of a rectangular area of length $L$ and width $W$. Within this area, $N_{TRP}$ TRPs, each of height $h_{TRP}$ are arranged in an equally spaced grid of $M$ rows and $N$ columns. The TRPs are spaced $D$ meters apart. UEs (such as autonomous factory robots) with antenna height $h_{UE}$ are dropped uniformly within the convex hull of the TRPs. Figure~\ref{fig:Example InF-DH Layout} shows an example of an InF-DH layout. In this paper, as specified in~\cite{3gpp-38-901} we use $L = 120 \text{m}, W = 60\text{m}, N_{TRP} = 18, h_{TRP} = 8\text{m}, M = 3, N = 6, D = 20\text{m}, \text{ and } h_{UE} = 1.5\text{m}$. 
\subsection{Channel Model for InF-DH}
Here, we describe the InF-DH channel model, as defined in~\cite{3gpp-38-901}, including:
\begin{itemize}
    \item Clutter (multipath) and its resulting fading and shadowing effects
    \item Path Loss model
    \item Signal model, channel estimation and channel measurements 
\end{itemize}

\subsubsection{Clutter Parameters}
In indoor factories, clutter objects such as autonomous robots and other machinery may result in NLoS dominant channel conditions and shadow fading (defined by standard deviation $\sigma_{SF}$). The InF-DH scenario is characterized by a clutter density $r$ (which represents the percentage of floor area covered by clutter objects), clutter height $h_{c}$ and clutter size $d_{c}$. In this paper, we use $\sigma_{SF} = 4$, $r = 60\%$, $h_{c} = 6m$ and $d_{c} = 2m$

\subsubsection{Path Loss model}
The Path Loss model for InF-DH is given by the following equations:

\begin{align}
PL_{NLoS} &= 33.63 + 21.9 \log_{10}(d_{3D}) + 20 \log_{10}(f_c) \label{eq:general_path_Loss} \\
PL_{LoS} &= 31.84 + 21.5 \log_{10}(d_{3D}) + 19 \log_{10}(f_c), \label{eq:LoS_path_Loss}
\end{align}
where $d_{3D}$ is the 3D distance between the transmitter and receiver antennas in meters, and $f_c$ is the carrier frequency in GHz. In this paper, we use $f_{c} = 3.64GHz$. The overall Path Loss in dB, in the InF-DH scenario is calculated as:
\begin{align}
PL = \max\left(PL_{NLoS}, PL_{LoS}\right)
\end{align}

\subsection{Signal model and channel estimation} \label{Ch Est}
Consider an OFDM system with bandwidth $B$ and $N$ sub-carriers. The received signal at the UE, in the $k^{th}$ sub-carrier is given by,
\begin{align}
Y(k) = \sqrt{\frac{P_{t}PL}{N}}H(k)X(k) + W(k),
\end{align}
where $P_{t}$ is the transmit power, $PL$ is the Path Loss, $H(k)$, $X(k)$ and $W(k)$ are the channel coefficient, the transmitted reference signal (for example, Downlink Positioning Reference Signal in 5G NR) and the receiver noise, respectively, at the $k^{th}$ sub-carrier. The estimated channel $\Tilde{H}(k)$, at the $k^{th}$ sub-carrier, can be obtained using standard methods such as $\Tilde{H}(k) = Y(k)/X(k)$.

In this paper, in lieu of actual reference signal transmission and channel estimation, we use a channel modeling tool called QuaDRiGa~\cite{quadriga}, that generates time and frequency domain channel coefficients for specific 3GPP channel scenarios, including InF-DH. For the given bandwidth $B$, subcarrier spacing $\Delta_{f}$ and all other layout parameters described above, QuaDRiGa generates a length $N$ frequency domain channel. From this, we obtain the time domain channel (CIR) by $N$-point IFFT which we further truncate to the first $L$ coefficients. The truncated CIR is denoted by $\tilde{h}(n)$ where $n = 1, 2, \dots L$. In this paper we use, $B = 100\text{ MHz}$, $N = 4096$, $\Delta_{f} = 30 \text{ kHz}$ and $L = 256$ taps. For the AI/ML model input, we use $\tilde{h}(n)$ normalized by the transmit power (assumed to be $24dBm$ for each TRP). 

In addition to the CIR, we also use RSRP measurements which are obtained as follows, 
\begin{align}
\text{RSRP} = \frac{1}{L} \sum_{n=1}^{L} |\tilde{h}(n)|^2
\label{RSRP}
\end{align}

\section{Proposed method for AI/ML-based positioning} \label{LocNet Model}
In this paper, we adopt a supervised learning approach in which a neural network is trained to predict the position using CIR and RSRP as inputs. This section describes the dataset generation, model architecture and training procedures used in this paper.


\subsection{Dataset Generation} \label{sec: datasets}
Three types of datasets are generated. These are shown in Table~\ref{tab:dataset_table}. In all three, the ground truth label (model output) is the UE's $(x, y)$ position. The difference is in the types of model input combinations. 

\subsubsection{CIR} This is a baseline dataset that consists only of CIR. Each data instance is of size $N_{TRP} \times L$.

\subsubsection{CIR + RSRP} In addition to CIR, RSRP values can also be included in the model input. This is motivated by the fact that higher values of RSRP would indicate that the measurements from that corresponding TRP are an important component in estimating the UE's location. For each data instance, the RSRP values from all the TRPs are interleaved with the CIRs. The same RSRP value is repeated in the real and imaginary dimensions. Consequently, each data instance is of size $2N_{TRP} \times L \times 2$. 

\subsubsection{CIR + RSRP + TRP Ratio} This dataset is the same as the CIR + RSRP dataset except for the third dimension, in which, in addition to the real and imaginary parts, we add another $N_{TRP}\times L$ matrix. Each element of this matrix holds the same value i.e., the TRP ratio. We define the TRP ratio as the ratio of number of TRPs from which measurements are available ($N^{'}_{TRP}$) to the total number of TRPs ($N_{TRP})$. Each data instance is of size $2N_{TRP} \times L \times 3$.

\newcolumntype{C}[1]{>{\centering\arraybackslash}m{#1}}
\begin{table*}[h!]
\centering
\renewcommand{\arraystretch}{2}
\caption{Generated dataset types, shapes, and descriptions}
\begin{tabular}{|C{3.5cm}|C{3cm}|C{2cm}|C{6cm}|} 
\hline
\textbf{Dataset Type} & \textbf{Model Input Size} & \textbf{Model Output Size} & \textbf{Description} \\ 
\hline
CIR & $N_{TRP} \times L \times 2$ & $2$ & Only Channel Impulse Response\\ 
\hline
CIR + RSRP & $2N_{TRP} \times L \times 2$ & $2$ & CIR and RSRP interleaved in the second dimension. Same RSRP value repeats across real and imaginary \\ 
\hline
CIR + RSRP + TRP Ratio & $2N_{TRP} \times L \times 3$ & $2$ & CIR and RSRP interleaved in the second dimension. Same RSRP value repeats across real and imaginary. TRP Ratio in the third dimension, repeats $2N_{TRP}\times L$ times  \\ 
\hline
\end{tabular}
\label{tab:dataset_table}
\end{table*}

\subsection{AI/ML Model Architecture}
\begin{figure}[ht!]
    \centering
        \includegraphics[width=0.485\textwidth]{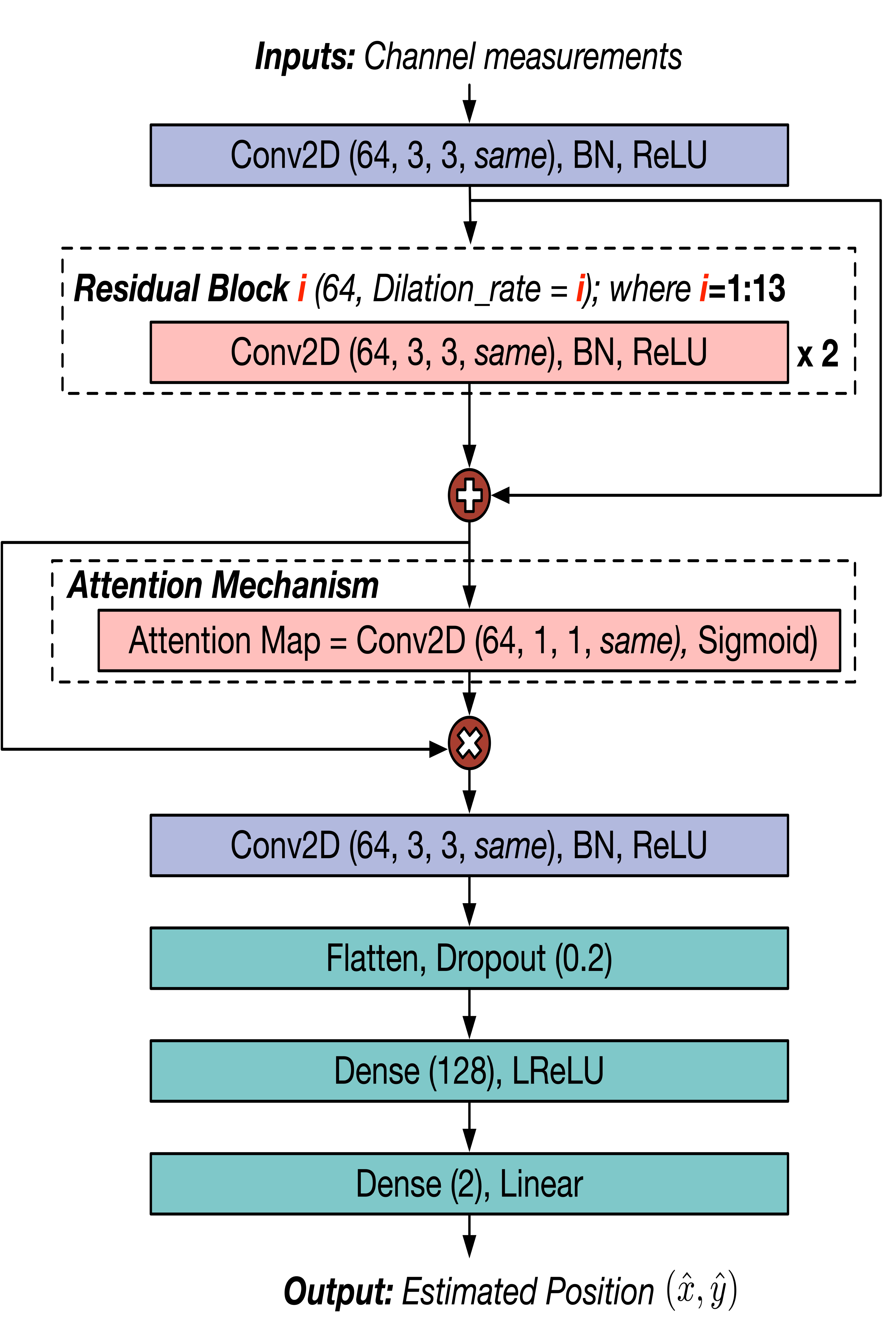}
    \caption {Proposed LocNet model architecture}
  \label{fig: locnet_arch}
\end{figure}

In this paper we develop a neural network, which we term LocNet. The IQ samples in the datasets used in this paper are analogous to pixel values in an image. Hence, given the similarity in input structure, neural network architectures that work well on images, could work even for position estimation. As a result, the architecture of LocNet, shown in Figure~\ref{fig: locnet_arch}, consists of a ResNet-like backbone coupled with dilated convolution and an attention mechanism. 

LocNet processes the input though an initial 2D convolution layer followed by repeated residual blocks. The residual blocks also contain 2D convolutional layers with dilated convolution operations to improve the receptive field of the network. LocNet has 13 residual blocks with two convolutional layers in each block. A shortcut connection from the input of the first residual block and the output of the last residual block allows information to propagate directly between the deep and shallow layers. 

The next part of the neural network is what we term an Attention Mechanism which consists of an attention layer followed by a multiplication operation. The attention layer is a 2D convolutional layer with a sigmoid activation function. The output of this layer, which we call an attention map, is used as a filter with which the input to the attention layer is multiplied, element-wise. The sigmoid activation function restricts the values in the attention map to the interval [0, 1]. As the model learns, the values in the attention map that are closer to 1 help highlight the more important features in the data. 

The output of the Attention Mechanism passes through another 2D convolutional layer, before being flattened and passed through a dense layer, followed by the output (predicted position).  Between the flatten and dense layers, we use dropout for regularization purposes. Further details on the kernel sizes, dilation rates, batch normalization and activation functions can be found in Figure~\ref{fig: locnet_arch}. 

The model is trained using the loss function defined as $Loss = \frac{1}{2} \sqrt{(\hat{x} - x)^2 + (\hat{y} - y)^2}$ where $(\hat{x}, \hat{y})$ and $(x, y)$ represent the predicted and true positions respectively. 

\section{Simulation Results}
In this section we present simulation results showing the positioning accuracy performance of the LocNet model. In this paper, we characterize positioning accuracy by the 90th percentile point on the Cumulative Distribution Function (CDF) curve of positioning accuracy in the horizontal direction. In all simulations, we use randomly shuffled datasets of size 80000 samples from which we use a train-test split of $80-20\%$. A further $20\%$ of the training data is used for validation. 

\subsection{Choice of model input}
\begin{figure}[htpb]
    \centering
    \includegraphics[width=0.485\textwidth]{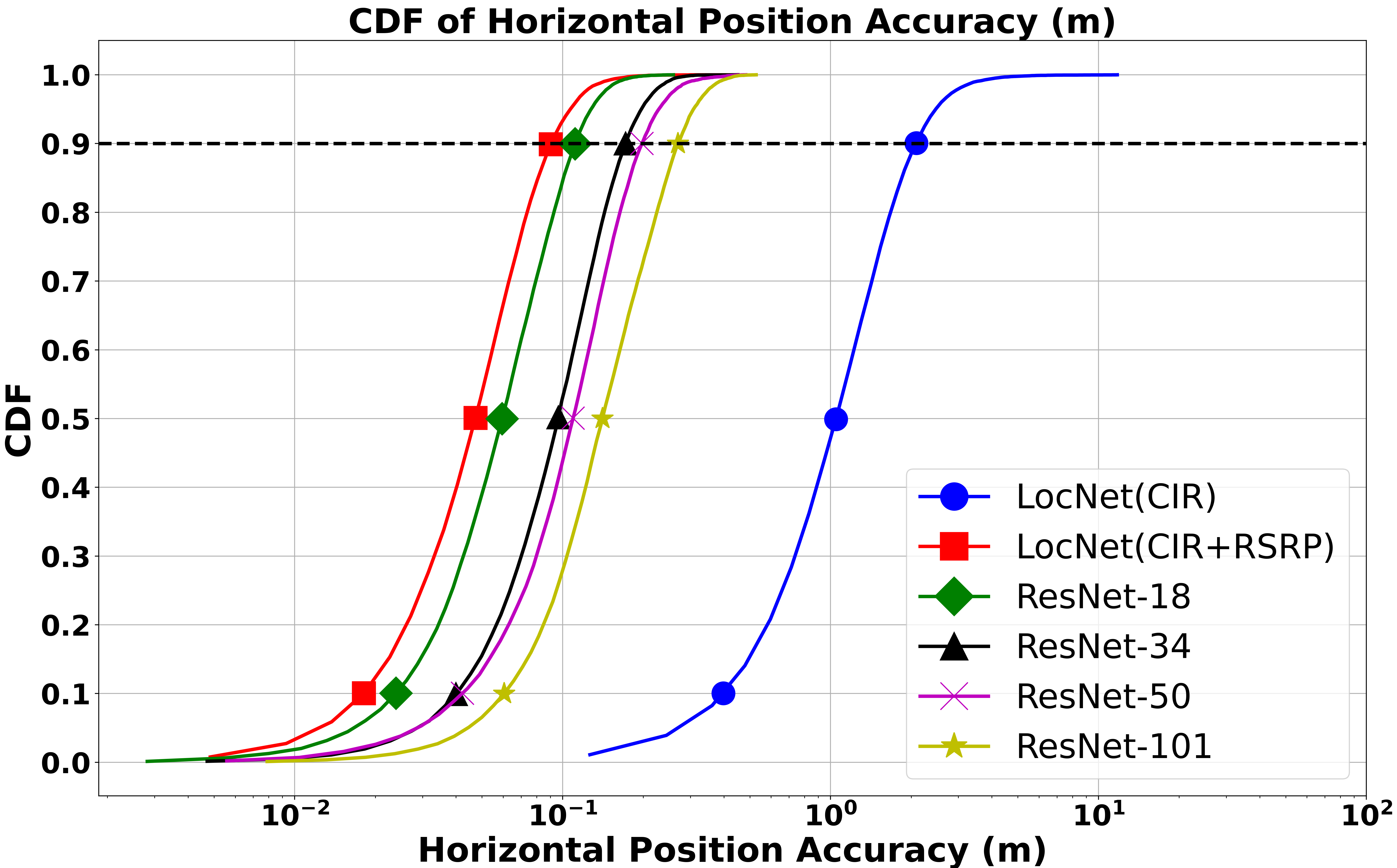}
    \caption{Comparison of 90th percentile Horizontal positioning accuracies (test dataset) of LocNet (trained with CIR and CIR+RSRP datasets) versus ResNet variants (ResNet-18, 34, 50, 101) (trained on the CIR+RSRP dataset)}
    \label{fig: locnet_vs_resnet}
\end{figure}

Figure~\ref{fig: locnet_vs_resnet} shows CDF curves of the horizontal positioning accuracy. It can be seen that LocNet with both CIR and RSRP as model input significantly outperforms the case with only CIR as input. The improved performance can be attributed to the model learning to give more weight to CIRs corresponding to higher RSRPs versus those corresponding to lower RSRPs. These curves represent the best case scenario in which all $18$ TRPs provide measurements. 

Furthermore, LocNet with CIR + RSRP input shows better positioning accuracy compared to other commonly used ResNet models, showing that our addition of both the dilation and Attention Mechanisms has a positive impact on model performance. 

Table~\ref{tab: locnet_vs_resnet_complxity} shows the 90th percentile values of the horizontal positioning accuracy for LocNet, ResNet-18, ResNet-34, ResNet-50 and ResNet-101. It can be seen that LocNet achieves a positioning accuracy close to $9\text{cm}$. We note that though the other ResNet architectures also achieve centimeter-level positioning accuracy, their complexities (number of parameters) are much higher. For example, we observe that ResNet-18 achieves a positioning accuracy of almost $11\text{cm}$ but at almost $4\times$ the model complexity. Also, in the case of conventional ResNets, increasing the number of residual blocks leads to poor positioning accuracies due to overfitting. 

\begin{table}[htpb]
\centering
\renewcommand{\arraystretch}{2} 
\caption{Comparison of model complexity and positioning accuracy (test dataset) of LocNet versus traditional ResNet models trained with the CIR+RSRP dataset}
\label{tab: locnet_vs_resnet_complxity}
\begin{tabular}{|c|c|c|}
    \hline
    \textbf{Model} & \textbf{Params (M)} & \textbf{90\% Horizontal Pos. Accuracy (m)} \\
    \hline
    LocNet & 2.9 & 0.089\\
    \hline
    ResNet-18 & 11.7 & 0.111\\
    \hline
    ResNet-34 & 21.8 & 0.172\\
    \hline
    ResNet-50 & 25.6 & 0.198\\
    \hline
    ResNet-101 & 44.06 & 0.27\\
    \hline
\end{tabular}
\end{table}

\subsection{Variable TRPs providing measurements}
In practical deployments it may not be feasible for a UE to obtain measurements from all $18$ TRPs. Network schedulers may dynamically optimize radio resources leading to a variable number of measurements that a UE can obtain i.e., the value of $N^{'}_{TRP}$ may constantly be changing. It is computationally prohibitive to train separate models for each of the $\binom{18}{N^{'}_{TRP}}$ possible combinations. So, we retrain the LocNet architecture such that the same model can predict the position from a variable number measurements. 

Each of the three datasets described in Section~\ref{sec: datasets} are modified as follows. Let us assume that each data instance represents a channel measurement made by a UE at a certain location. For a given UE and for a given value of $N^{'}_{TRP}$ we select the TRPs that have the top $N^{'}_{TRP}$ RSRP values. The CIR values corresponding to the TRPs not providing measurements are replaced by $0$. The RSRP values are replaced by a very small value ($-500\text{dBm}$). 

\begin{figure}[h!]
    \centering
    \includegraphics[width=0.485\textwidth]{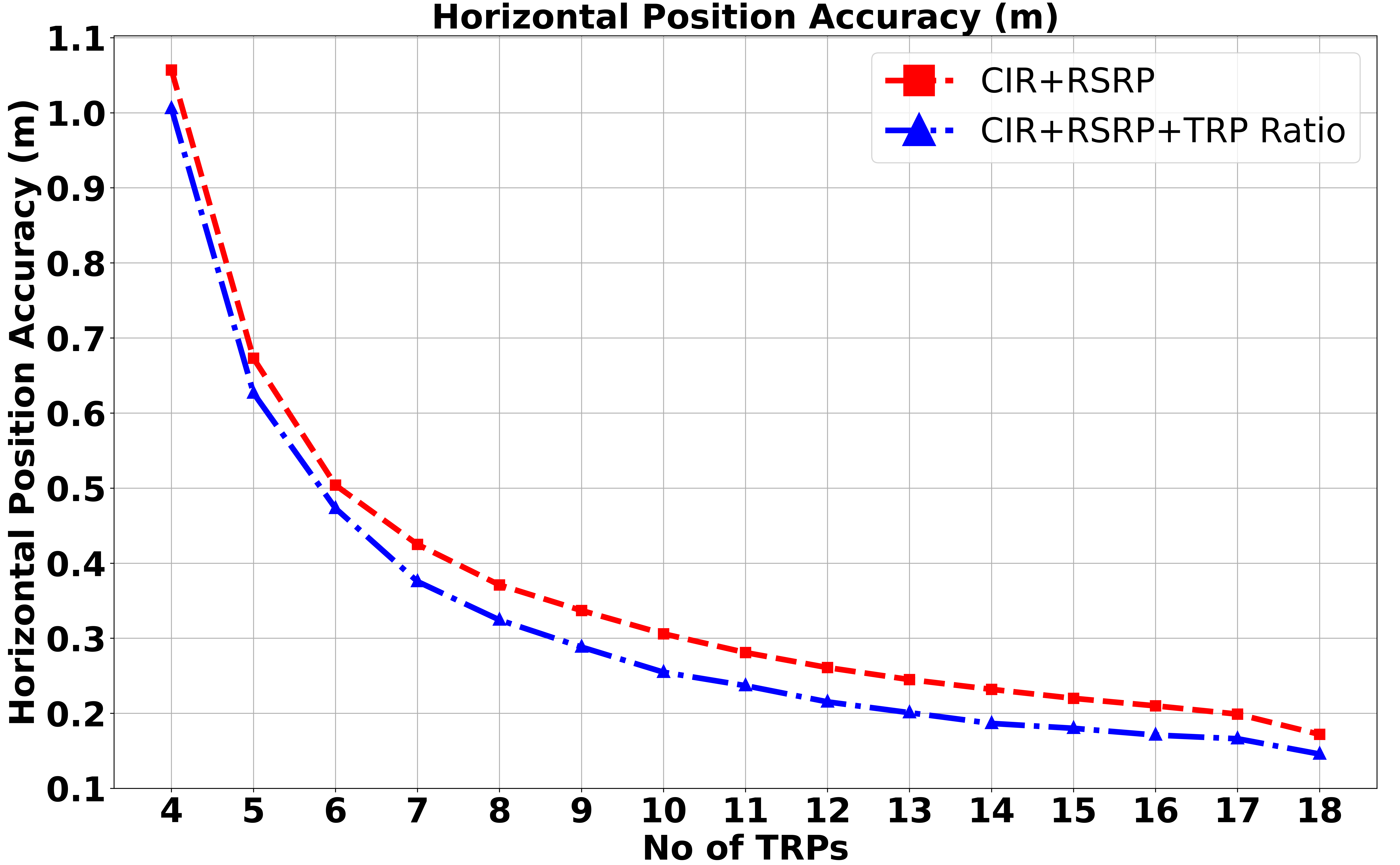}
    \caption{90th percentile horizontal positioning accuracy (test dataset) for different number TRPs. There is an improvement in positioning accuracy due to the inclusion of TRP Ratio in the model input.}
    \label{fig: variable_trp}
\end{figure}

Similar to Figure~\ref{fig: locnet_vs_resnet}, Table~\ref{tab: variable_trp} also shows that for all values of $N^{'}_{TRP}$, the dataset containing CIR only performs worse than the case with CIR+RSRP. In the case of variable TRPs providing inputs, there are further gains when the TRP Ratio is also included. These gains are shown in Figure~\ref{fig: variable_trp}. Both CIR+RSRP and CIR+RSRP+TRP Ratio achieve centimeter-level accuracy with measurements from even as low as 5 TRPs. But there is approximately $15\%$ improvement when the TRP Ratio is also provided. This could be significant in some Indoor Factory applications requiring precise robotic manoeuvres. 

For Figure~\ref{fig: variable_trp}, 35000 data samples correspond to $N^{'}_{TRP}$ values of $4$ to $17$ in equal proportions. The remaining 45000 samples correspond to measurements from all $18$ TRPs. The resultant dataset of 80000 samples is then split for training, validation and testing as described above. 

\begin{table}[htpb]
\renewcommand{\arraystretch}{2}
\centering
\caption{90th percentile horizontal positioning accuracy (test dataset) for different number of TRPs and different types of datasets}
\label{tab: variable_trp}
\begin{tabular}{|C{1cm}|C{1.5cm}|C{1.5cm}|C{2cm}|} 
\hline
\textbf{No. of TRPs} & \textbf{CIR Dataset} & \textbf{CIR+RSRP Dataset} & \textbf{CIR+RSRP+TRP Ratio Dataset} \\
\hline
4  & 4.478 &1.057  & 1.0057 \\
\hline
5  & 3.796 &0.673  & 0.6266 \\
\hline
6  & 3.379 &0.504  & 0.4731 \\
\hline
7  &3.138  & 0.425 & 0.3756 \\
\hline
8  &2.990  &0.371  & 0.3244 \\
\hline
9  &2.945  &0.337  & 0.2883 \\
\hline
10  &2.862  &0.306  & 0.2548 \\
\hline
11  &2.762  &0.281  & 0.2369 \\
\hline
12  &2.626  &0.261  & 0.2153 \\
\hline
13  &2.530  &0.245  & 0.2009 \\
\hline
14  & 2.404 &0.232  & 0.1866 \\
\hline
15  & 2.296 &0.220  & 0.1801 \\
\hline
16  & 2.153 &0.210  & 0.1711 \\
\hline
17  & 2.031 &0.199  & 0.1662 \\
\hline
18  & 1.795 &0.172  & 0.1458 \\
\hline
\end{tabular}
\end{table}

\subsection{Impact of label error}
To emulate real-world data collections scenario, we analyze the model performance with erroneous ground truth (position) labels. We use the CIR+RSRP dataset with $N_{TRP} = 18$. Error added to the ground truth labels is sampled from a truncated Gaussian distribution of 0 mean and standard deviation $\sigma$. The truncation is in the interval $[-2\sigma, 2\sigma]$. In this paper we consider $\sigma = 0.1m, 0.3m, 0.5m, 0.7m, 1m$. The dataset is constructed such that there are 16000 samples corresponding to each value of $\sigma$. After training, the model is evaluated on clean data consisting of the original ground truth labels. The model achieved a positioning accuracy of $0.153 m$ on the clean data, demonstrating that the proposed LocNet is robust to label error within the considered error limits.

\section{Conclusion and Future Work}
In this paper, we have developed an AI/ML model that achieves centimeter-level positioning accuracies in Indoor Factory scenarios. We have investigated the model performance for a variable number of TRPs that provide measurements. We have also studied the impact of ground truth labelling error on the robustness of the model. Future work could include generalizing the model across different drops of UE positions and channel scenarios. The impact of CIR length on model performance could also be studied. 

\section*{Acknowledgement}
The authors would like to thank the Indian Department of Telecommunications (DOT) for their financial support of the Indigenous 5G Testbed initiative and the Ministry of Electronics and Information Technology (MeitY) for their funding of this research through the project "Next Generation Wireless Research and Standardization on 5G and Beyond".

\bibliographystyle{IEEEtran}
\bibliography{pos_est}
\end{document}